\documentclass[referee]{gji}
\usepackage{timet}
\usepackage{amssymb}

\title[CMB and $J_2$ perturbations from 2004 Sumatra earthquake]
  {Core-mantle boundary deformations and $J_2$ variations resulting from the 2004 Sumatra earthquake}
\author[V. Cannelli et al.]
  {V. Cannelli, D. Melini, P. De Michelis, A. Piersanti, F. Florindo \\
  Istituto Nazionale di Geofisica e Vulcanologia, Via di Vigna
Murata 605, I-00143 Rome, Italy }
\date{Received xxxx; in original form xxxx}
\pagerange{\pageref{firstpage}--\pageref{lastpage}}
\volume{200}
\pubyear{1998}

\begin{document}

\label{firstpage}

\maketitle

\begin{summary}

The deformation at the core-mantle boundary produced by the 2004 Sumatra earthquake is investigated by means of a semi-analytic theoretical model of global coseismic and postseismic deformation, predicting a millimetric coseismic perturbation over a large portion of the core-mantle boundary.
Spectral features of such deformations are analysed and discussed.
The time-dependent postseismic evolution of the elliptical part of the gravity field ($J_2$) is also computed for different asthenosphere viscosity models.
Our results show that, for asthenospheric viscosities smaller than $10^{18}$ Pa s, the postseismic $J_{2}$ variation in the next years is expected to leave a detectable signal in geodetic observations.

\end{summary}


\section{Introduction}

The devastating megathrust earthquake occurred on December 26th, 2004
off the west coast of northern Sumatra was the second-greatest event
ever registered, according to current estimates which give a moment
magnitude $M_w=9.3$. Such an exceptional event produced measurable
effects on many geophysical observables. It has
been shown \cite{sumatraosc} that the 2004 Sumatra earthquake was able to excite Earth's free oscillations of exceptionally large amplitude.
These oscillations, characterized by periods $T>1000$ s, remained
observable for weeks in broadband seismic data providing
information on the size and duration of the event itself. At the same
time, this event is expected to have produced a jump in the rotational pole\'s secular motion \cite{eosrotazione,gross}, yet current geodetic measurements were not able to detect it, probably because it is shadowed by other effects of atmospheric and oceanic origin. Associated static surface deformation field data shows \cite{sumatragps,boschi2006} that the Sumatra
earthquake produced static offsets of the order of 1 mm recorded by
continuous GPS stations located up to 5000 km away from the
epicentre. On the basis of the aforementioned points, it is reasonable
to expect that the static deformation associated with the event could
have affected also a large part of the Earth's interior.

The main aim of this work is to contribute to the characterization of the
global effects of Sumatra event on core-mantle boundary (CMB) topography and on the $J_2$ gravitational field coefficient. Using a semi-analytical deformation model,
we estimated the amplitude and shape of the CMB topography changes as the result of this event both in the purely elastic ($t=0$) and fluid ($t\to \infty$) limits.
In addition, we have performed a spherical harmonic decomposition of the coseismic CMB deformation field, investigating its symmetric properties. This analysis has been motivated by the possible connection between CMB deformations with axial and equatorial symmetry and core flow perturbations \cite{dumberry1}.
As a result, we have found that the axial and equatorial symmetric component of the CMB deformation has an
amplitude of the order of a fraction of millimeter.
It is worth noting that, although the effect of Sumatra earthquake on the CMB may appear modest, 
the amplitude of this deformation turns out to be comparable with the distortion of the elliptical surfaces of constant density at the CMB surface resulting from torsional oscillations in the core \cite{dumberry1}.
While this evidence alone does not imply a causal relationship between the coseismic deformation field resulting from giant earthquakes and core flow perturbations, our results suggest that the CMB deformation of seismic origin has the potential to interfere with core dynamics. For instance, it has been recently a matter of debate whether a seismic perturbation of the CMB could trigger a flow instability, leading to a geomagnetic jerk \cite{eosflorindo,commentdumberry,replydumberry}; in this case, a jerk should follow the seismic event, after a suitable time delay to allow for the signal to propagate through the weakly conducting mantle.

Following the evidence that the main contribution to the deformation field comes from the lowest degrees coefficients of the spherical harmonic expansion, we have investigated the detailed time-dependent evolution of the perturbation to the elliptical part of the gravity field, $J_2$. Our results show that the Sumatra event produces a negative variation of $J_2$, confirming the well-known tendency of earthquakes to reduce the Earth's oblateness \cite{cg2,alfonsi98}; this effect turns out to be further enhanced by the postseismic relaxation, according to our model.

A sensitivity analysis, performed with varying asthenosphere viscosities, shows that, for values smaller than $10^{18}$ Pa s, the postseismic effect on $J_2$ remains comparable with the main secular trend for several years after the event. In the next years the analysis of available $\dot{J}_2$ data would allow us to put a lower bound to asthenospheric viscosity, which is still highly controversial issue \cite{pollo,piersanti99,marquart}.

\section{Modeling approach}

To compute the coseismic and postseismic effects of the Sumatra earthquake on the core-mantle boundary, we used the semi-analytical model originally proposed by Piersanti et al. \shortcite{piersanti95}, which is a spherical, self-gravitating, incompressible model with Maxwell viscoelastic rheology.

This model computes physical observables on the Earth's surface; however, its formulation straightforwardly allows to extract the deformation field at the CMB. In fact, the harmonic components of the physical quantities at the CMB (deformation, geopotential and stress tensor) are imposed as boundary conditions through the so-called ``continuity matrix'', so that the spheroidal and toroidal parts of the solution at the CMB ($r=r_c$) can be written as follows:

$$ \mathbf{y}(r_c) = \mathbf{I}_s(r_c) \mathbf{c}_c $$
$$ \mathbf{z}(r_c) = \mathbf{I}_t(r_c) c_c $$

where $\mathbf{y}$ is a 6-vector corresponding to the spheroidal part of the problem, $\mathbf{z}$ is a 2-vector corresponding to the toroidal part, $\mathbf{I}_s$ and $\mathbf{I}_t$ are the spheroidal and toroidal continuity matrices, whose expressions are given by Sabadini et al. \shortcite{cmbj} and Piersanti et al. \shortcite{piersanti95}, and $\mathbf{c}_c$ and $c_c$ are, respectively, a vector and scalar constant to be determined by imposing traction-free boundary conditions at the Earth's surface.

The perturbation to the gravity field elliptical term $J_2$ is related by definition to the $l=2$, $m=0$ component of the geopotential as follows \cite{lambeck}: 

$$ \Delta J_2 = \frac{R_{T}}{GM_{T}} \phi_{2,0}(R_{T}) $$

where $R_{T}$ and $M_{T}$ are radius and mass of the Earth respectively, $G$ is the gravitational constant and $\phi_{2,0}$ is the second-degree harmonic coefficient of the perturbation to the gravitational potential.

While the computation of $\Delta J_2$ involves the $l=2$ harmonic term only, the evaluation of the deformation field requires the summation of hundreds of harmonic terms to gain a stable convergence. The CPU time needed to compute a single harmonic term increases strongly with the number of layers in the model \cite{boschi2000}; in order to be able to employ a realistic, refined stratification and at the same time keep the computation time within reasonable limits, we adapted the analytical model formulation to the purely elastic and fluid cases by taking the limits $t\to 0$ and $t\to\infty$ respectively, which in the Laplace domain correspond to $s\to\infty$ and $s\to 0$. 
In this way, we compute the full CMB deformation field in the elastic and fluid limits and give the transient postseismic evolution of $\Delta J_2$, which corresponds to Earth oblateness and is an indicator of the ellipticity evolution.

The stratification model used in our computations is built by 
adopting the PREM (Preliminary Reference Earth
Model) \cite{prem} for mantle and crust and a uniform fluid core,
with rigidity $\mu_c = 0$ and density $\rho_c = 10.93 \mbox{ kg/m}^3$, obtained by volume-averaging PREM core layers. 
The viscosity of the layers has been assigned by interpolating the viscosity model given by Mitrovica \& Forte \shortcite{forte}. The resulting model has a total of 43 homogeneous layers and a uniform fluid core; its density, rigidity and viscosity profiles in mantle and crust are represented in figure \ref{stratif}. It is to note that, while the PREM model is compressible, the analytical formulation of our model is based on an incompressible rheology; that is to say, we adopt a modified version of PREM with $\lambda\to\infty$. This approximation certainly affects our results, as discussed in detail by Nostro et al. \shortcite{nostro}, but presently it is an unavoidable choice if we want to take into account simultaneously viscoleasticity, self-gravitation and sphericity.

The seismic source has been modeled using the five point sources obtained by Tsai et al. \shortcite{multipleCMT}. These seismic sources have been computed by fitting with the CMT method \cite{CMT} the long-period seismograms from the IRIS Global Seismographic Network and account for a cumulative energy release corresponding to $M_w=9.3$.

\section{Perturbation of the core-mantle boundary}

In what follows we show and discuss the results we obtained in relation to CMB effects of the Sumatra earthquake.
Figures \ref{componenti} and \ref{arrowhoriz} show, respectively, the scalar components of the dislocation vector
$\mathbf{u} = (u_r,u_\theta,u_\phi)$ and the horizontal displacements at the CMB in an orthographic projection,
centered on the location of the composite CMT source obtained by Tsai et al. \shortcite{multipleCMT}, in the coseismic case.

In particular, figure \ref{componenti} shows the horizontal components along colatitude and longitude directions ($u_{\theta}$, $u_{\phi}$), the radial ($u_{r}$) component and the absolute value of the displacement (${\mid}\mathbf{u}{\mid}$); the radial component shows a local CMB depression of about $4$ mm by the event location. Starting at distances of $50^\circ$ from the source location we observe a global vertical displacement of about $0.5$ mm. 

Figure \ref{arrowhoriz} shows the horizontal displacements evaluated at CMB. The whole CMB surface is affected by appreciable displacements, with deformations still of the order of a fraction of millimeter even at extremely large epicentral distances. We also observe that the horizontal displacements are directed westward near the equator while near the poles the direction is opposite. 
The radial and horizontal displacement at the CMB computed by our numerical method are of the same order of magnitude.

A spectral harmonic analysis has been performed in order to better understand the symmetric properties of the observed field deformations. Indeed, Dumberry \& Bloxham \shortcite{dumberry1} pointed out that only a CMB deformation satisfying axial and equatorial symmetry has the potential to interact with fluid core flows, possibly triggering a flow instability.

Because of the location and north-south orientation of the fault plane, the CMB deformation field is approximately symmetric with respect to the equator plane (see figures \ref{componenti} and \ref{arrowhoriz}). On the other hand, the rupture geometry is not axisymmetric, so the deformation field has no intrinsic axial symmetry. Nevertheless, if we write the deformation field as a sum of spherical harmonic terms, we can extract the axial and equatorial symmetric terms and evaluate their amplitude.

Let us now write the deformation field $\mathbf{u}(\theta,\phi)$ as a
sum of spherical harmonic functions:
\begin{equation}
\label{arm1}
\mathbf{u}(\theta,\phi) = \sum_{l=0}^\infty \sum_{m=-l}^l \mathbf{c}_{lm} Y_{lm}(\theta,\phi)
\end{equation}
with $\mathbf{c}_{lm} = ( c_{lm}^{(r)}, c_{lm}^{(\theta)},
c_{lm}^{(\phi)})$ being the vector whose elements are the harmonic coefficients of the expansion of the deformation along $\hat{r}$, $\hat{\theta}$ and $\hat{\phi}$ directions, respectively. The spherical harmonic functions $Y_{lm}$ are defined as:
\begin{equation}
\label{arm2}
Y_{lm}(\theta,\phi) = \sqrt{ \frac{2l+1}{4\pi} \frac{(l-m)!}{(l+m)!} } e^{im\phi} P_{lm}(\cos\theta)
\end{equation}
with $P_{lm}$ being the associated Legendre functions. The spherical
harmonics satisfy axial and equatorial symmetry only for even $l$ and
$m=0$, so we can write the symmetric component of the deformation field
as:
\begin{equation}
\label{arm3}
\mathbf{u}_S(\theta) = \sum_{l\: even} \mathbf{c}_{l0} Y_{l0}(\theta)
\end{equation}
where we dropped the $\phi$ dependence on the $m=0$ spherical harmonic functions.
The harmonic coefficients $c_{l0}$, because of the orthonormality
properties, can be immediately evaluated by:
\begin{equation}
\label{arm4}
\mathbf{c}_{l0} = \int \mathbf{u}(\theta,\phi) Y_{l0}(\theta) d\Omega
\end{equation}
with $d\Omega = \sin\theta d\theta d\phi$ being the solid angle element.
In figure \ref{symm1} we plot the harmonic coefficients
$\mathbf{c}_{l0}$ for $l=0,2,...,20$, computed by numerically integrating equation (\ref{arm4}). The harmonic
amplitudes show that a non-negligible amount of deformation associated with the lowest degrees
satisfies the symmetry requirements. From figure \ref{symm1} we see that the main contribution to symmetric term $\mathbf{u}_S$ is given by the $u_\phi$ component of the CMB deformation.

In figure \ref{symm2}, we show the symmetric term of the deformation $\mathbf{u}_S$, as in equation (\ref{arm3}), and the
associated residual $\mathbf{u}_R = \mathbf{u} -
\mathbf{u}_S$. Since the symmetric components of the deformation field are associated with lowest harmonic degrees, as shown in figure \ref{symm1}, they have been computed using only the spectral components $l\le 20$. From figure
\ref{symm2} we see that the CMB deformation field exhibits spectral
components that satisfy axial and equatorial symmetry accounting for a
considerable part of the total deformation. For example, the component whose
symmetric part has the largest amplitude is $u_\phi$, which has a peak
in the equatorial zone reaching $0.8$ mm. Also the radial deformation,
$u_r$, shows a non-negligible symmetric component, with a considerable
range of latitudes where the deformation amplitude exceeds $0.1$ mm,
while the component $u_\theta$ turns out to have the smallest symmetric
term.

We verified that the CMB deformation is strongly dependent from the dip angle of the seismic source: smaller dip angles result in less pronounced deformation effects on the CMB. This can be qualitatively explained by noting that in a source mechanism with a small dip angle there is little amount of slip in the radial direction, which is the component on which the CMB effects are most dependent.

In figure \ref{fluid} we plot the full CMB deformation field in the fluid limit ($t\to\infty$). The horizontal components of the deformation field are greatly enhanced by the postseismic relaxation, while the radial component mean amplitude is comparable to the elastic case. As a result, the total deformation vector $\mathbf{u}$ is about an order of magnitude greater than the elastic case, with peak values of a few centimeters. These results can be interpreted as a direct consequence of the viscoelastic relaxation of a Maxwell body, which in the fluid limit cannot sustain tangential stresses; therefore, the major effects are expected on the horizontal components of the deformation.
Moreover, from a comparison of figures \ref{componenti} and \ref{fluid}, we see that the deformation field in the fluid limit has a smoother spatial variation than the elastic limit, so we expect a further redistribution of the harmonic components towards lower wavelengths.

\section{Coseismic and postseismic effects on $J_2$}

In this section we show the coseismic and postseismic effects of the Sumatra earthquake on the oblateness $J_2$ of the gravitational potential. This quantity is directly related to the Earth flattening $f=(a-c)/c$, where $c$ and $a$ are the axial and equatorial radius respectively, so that positive variations of $J_2$ corresponds to an increase in the Earth's oblateness; for an homogeneous sphere, the simple relation $J_2=2f/5$ holds. 

In figure \ref{j2} we show the long-term time dependence of $J_2$ resulting from the mass redistribution following the Sumatra earthquake. It was computed adopting the stratification model shown in figure \ref{stratif}. We see that, similarly to most thrust subduction earthquakes, the Sumatra event gives a negative variation of $J_2$, corresponding to a decrease of the Earth oblateness \cite{alfonsi98}. This variation in the elastic limit is $\Delta J_2 = -0.30\times 10^{-10}$ and is further enhanced by the postseismic evolution, up to $\Delta J_2 = -1.8 \times 10^{-10}$ in the fluid limit. The transient evolution of $\Delta J_2$ exhibits the largest variation for $\Delta t \sim 10^2 \div 10^3$ yr. This feature is to be ascribed to the detailed viscosity structure of the model, since it is dependent on the complex convolution of the model relaxation times. Incidentally, this kind of time dependence can be found in the postseismic modeling of other observables, such as the deformation field \cite{nostro,boschi2000}.

We note that our result for the elastic limit is in good agreement with that by Gross \& Chao \shortcite{gross}, which estimated $\Delta J_2 = -0.24 \times 10^{-10}$ using the CMT source model obtained by Tsai et al. \shortcite{multipleCMT} and a PREM elastic stratification.
The coseismic variation of $J_2$ resulting from the Sumatra earthquake is therefore roughly equal to the variation occurring over a year due to the secular linear drift, which is $\dot{J}_2 \simeq -0.28 \times 10^{-10} \mbox{ yr}^{-1}$ \cite{j2errore,coxchao}, and it is two orders of magnitude greater than the mean annual $J_2$ variation associated with global seismic activity \cite{cg2,alfonsi98}.

This fact is particularly important since, as pointed out by Alfonsi \& Spada \shortcite{alfonsi98} the average effects of seismic activity and seismic tectonic movements tend to cancel each other both being of the order of $10^{-13} \mbox{ yr}^{-1}$ and with opposite sign. Only with an exceptional event like the Sumatra earthquake, we have the chance to register its effects on $J_2$.

While the global coseismic deformation produces a jump in the $J_2$ evolution, the postseismic relaxation of the ductile asthenospheric layers is expected to give a continuous temporal variation of $J_2$ that will be superimposed to its secular drift. In what follows, we have computed the short-timescale evolution of $J_2$ for various asthenosphere viscosities, to infer whether the viscoelastic relaxation may leave a detectable signature on the measured time-histories.

For this purpose a simplified three-layer stratification model was employed, with an 80 km elastic lithosphere, a 200 km asthenosphere with variable viscosity and a uniform mantle with a constant viscosity of $10^{21}$ Pa s.

In figure \ref{rateJ2} we show the time evolution of $\dot{J}_2$ over a period of twenty years for asthenosphere viscosities $\eta_{1}=10^{16}$, $\eta_{2}=10^{17}$ and $\eta_{3}=10^{18}$ Pa s. We see that low asthenospheric viscosities yield very large variation rates in the first years after the event, as a result of the low associated Maxwell times $\tau_i=\eta_i/\mu$, with $i=1,2,3$.

As such, these results can in principle be used to identify a lower limit for the asthenosphere viscosity on the basis of geodetic measurements of $J_2$. In fact, if we assume a likely detectability threshold for deviations of $\dot{J}_2$ from its secular drift and if no evidence of a such deviation is detected from available data, we can rule out the range of asthenosphere viscosities that produce perturbations on $\dot{J}_2$ above that threshold.

A reasonable value for the detectability threshold may be the associated formal error, which is about $10\%$ of the measured value \cite{j2errore}, i.e. $\sim 0.03 \times 10^{-10} yr^{-1}$. In figure \ref{rateJ2} the range of $\dot{J}_2$ values below that threshold is represented by shaded area; as can be seen, for asthenospheric viscosities $\eta = 10^{16}$ and $10^{17}$ Pa s, the effect of the Sumatra event on $\dot{J}_2$ would remain detectable for several years.

In table \ref{ratevar} the expected values of $\dot{J}_2$ in 2005 and 2006 for asthenosphere viscosities ranging from $10^{15}$ to $10^{22}$ Pa s are given. As can be seen, from viscosities up to $10^{17}$ Pa s we expect an evident signature in the data; viscosities greater than $10^{19}$ Pa s should not produce a detectable signal, while the signal associated with a viscosity of the order of $10^{18}$ Pa s lies marginal to the detectability threshold. We stress that this is just a general indication coming from a forward modeling; to apply this procedure to real data, a detailed sensitivity analysis will be needed.

\section {Conclusions}

In this work we have shown that even the core-mantle boundary is affected by a significant amount of seismic deformation produced by the giant Sumatra earthquake, with coseismic radial displacements of the order of a fraction of millimeter over the whole CMB surface and horizontal displacements even larger. By analyzing the spectral components of the coseismic deformation field at the CMB surface, we found that most of this deformation is associated with low degree harmonics, and that the deformation field has considerable spectral components characterized by axial and equatorial symmetry. These symmetric components account for a radial deformation of
the order of $0.1$ mm and horizontal deformation with peak values slightly less than a millimeter.

The CMB deformation field produced by the Sumatra earthquake turns out to be comparable to that resulting from core flow processes. In particular, we have verified that the CMB deformation is of the same order of magnitude of that resulting from core torsional oscillations \cite{dumberry1} and it is characterized by spectral components with similar symmetry. This suggests that the global deformation field from giant earthquakes has the potential to interfere with core processes. 
For instance, it has been recently suggested \cite{eosflorindo} that a perturbation of the CMB topography of seismic origin could, at least in principle, trigger a core flow instability which can lead to a geomagnetic jerk. While this possibility is rather controversial \cite{commentdumberry,replydumberry}, we have shown that the CMB deformation induced from a giant earthquake has a non-negligible amplitude and therefore should not be ruled out in playing a role in the triggering of core instabilities. In detail, Bloxham et al. \shortcite{bloxham2002} argued that torsional oscillations consistent with a geomagnetic jerk should have variations in amplitude of the order of 1 km/yr. Our results show that the component of the CMB deformation axisymmetric and symmetric about the equator has an amplitude of the order of a fraction of millimeter. Although the amplitude of this deformation is modest, it is important to take into account that inside the core a torsional oscillations flow of 1 km/yr results from a distorsion of the elliptical surface of constant density of 0.2 mm at the Earth surface and of about 0.15 mm at the CMB \cite{dumberry1}. This result opens the way to the sensibility that Sumatra event could really have triggered a jerk. A definitive answer on this issue would require a comprehensive modeling of core-mantle interaction, that is beyond the scope of the present work.

The postseismic evolution of the CMB deformation agrees with the known tendency of giant earthquakes to make the Earth rounder; that is,  the net effect of global seismicity is a decrease of $J_2$ over time \cite{alfonsi98}. In the fluid limit, only the horizontal components of the CMB deformation field are enhanced by about an order of magnitude, while the radial component remains of the same order of coseismic one; however, due to the extremely long time-scale of this processes, it is unlikely to expect any possible coupling with core flow geometry.

An important feature of the long wavelength deformation field associated with this event is the sensitivity to asthenospheric viscosity. For asthenosphere viscosity values smaller than $10^{18}$ Pa s, the postseismic $J_2$ variation lies above $10\%$ of the mean secular trend for several years after the earthquake and therefore is expected to be detectable in the $\dot{J}_2$ geodetic measurements. This would allow us, at least, to put a lower bound to mean asthenospheric viscosity that is still one of the outstanding issues in mantle rheology. In fact, when a detailed analysis of geodetic measurements of $J_2$ will become available, if a significant deviation of $\dot{J}_2$ from its secular trend will be evident, it will give an indirect estimate of mean asthenospheric viscosity; on the other hand, if no deviation will be evident, it indicates a lower limit of $\eta \gtrsim 10^{18}$ Pa s.

\begin{acknowledgments}
We thank the two anonymous reviewers for their helpful and incisive comments. This work was partly supported by a MIUR-FIRB research grant.
\end{acknowledgments}

\begin{table*}
\caption{Expected $J_2$ annual variation rate during 2005 and 2006 as a function of asthenosphere viscosity. A simplified three layered stratification model was used.}
\label{ratevar}
\begin{tabular} {@{}ccc}
\hline
Asthenosphere viscosity & $\dot{J}_2$ 2005        & $\dot{J}_2$ 2006 \\
     (Pa s)             & $(10^{-10}\cdot yr^{-1})$ & $(10^{-10}\cdot yr^{-1})$ \\
\hline
$10^{15}$ & $-1.753$ & $+0.031$ \\
$10^{16}$ & $-1.414$ & $-0.291$ \\
$10^{17}$ & $-0.256$ & $-0.224$ \\
$10^{18}$ & $-0.024$ & $-0.028$ \\
$10^{19}$ & $-0.003$ & $-0.003$ \\
$10^{20}$ & $-0.002$ & $-0.002$ \\
$10^{21}$ & $-0.002$ & $-0.002$ \\
$10^{22}$ & $-0.002$ & $-0.002$ \\
\hline
\end{tabular}
\end{table*}

% *********************************

\begin{figure*}
 \caption{Density, rigidity and viscosity profiles of the mantle and crust stratification model. 
          The model assumes a uniform fluid core with density $\rho_c = 10.93 \mbox{ kg/m}^3$, 
          obtained by volume-averaging the corresponding PREM layers.}
 \label{stratif}
\end{figure*}

\begin{figure*}
 \caption{CMB coseismic displacement vector components along radial ($u_r$), 
         colatitude ($u_\theta$) and longitude ($u_\phi$) directions and magnitude 
         $|\mathbf{u}|$.}
 \label{componenti}
\end{figure*}

\begin{figure*}
 \caption{CMB coseismic horizontal displacements resulting from the Sumatra earthquake. 
          The color scale represents the arrow length.}
 \label{arrowhoriz}
\end{figure*}

\begin{figure*}
 \caption{Harmonic coefficients of the component of the CMB deformation 
          field satisfying axial and equatorial symmetry.}
 \label{symm1}
\end{figure*}

\begin{figure*}
\caption{Spectral component $\mathbf{u}_S$ of the CMB deformation field 
satisfying both axial and equatorial symmetry and related residual 
$\mathbf{u}_R = \mathbf{u}-\mathbf{u}_S$.}
\label{symm2}
\end{figure*}

\begin{figure*}
 \caption{CMB displacement vector components in the fluid limit along radial 
         ($u_r$), colatitude ($u_\theta$) and longitude ($u_\phi$) directions 
         and magnitude $|\mathbf{u}|$.}
 \label{fluid}
\end{figure*}

\begin{figure*}
\caption{Time-dependent postseismic evolution of the perturbation to the elliptical part of the gravitational potential, $J_2$.}
\label{j2}
\end{figure*}

\begin{figure*}
 \caption{Time evolution of $J_2$ variation rate for a three layered stratification model. Different lines represent different values of asthenosphere viscosity. The shaded area corresponds to values of $\dot{J}_2$ below the detectability threshold ($\pm 0.03 \times 10^{-10} yr^{-1}$).}
 \label{rateJ2}
\end{figure*}

\label{lastpage}

\end{document}